\documentclass[%
aps, 
 prf,
 longbibliography,
preprint,
 amsmath,amssymb,
]{revtex4-1}

\usepackage{graphicx}
\usepackage{amsfonts}
\usepackage{amssymb}
\usepackage{float}
\usepackage{latexsym}
\usepackage{amsmath}
\usepackage{placeins}
\usepackage{subcaption}
\usepackage[euler]{textgreek}
\usepackage{textcomp}
\usepackage{siunitx}

\usepackage[table]{xcolor}

\newcommand{\eqr}[1]{(\ref{eq:#1})}

\newcommand{\eg}{\textit{e.g.}}
\newcommand{\bs}[1]{\boldsymbol{#1}}

\newcommand{\ie}{\textit{i.e.}}

\graphicspath{{./}}

\begin{document}

\title{Numerical study of natural oscillations of supported drops with free and pinned contact lines}

\author{Jordan Sakakeeny}
\author{Yue Ling}%
 \email{Stanley\_Ling@baylor.edu}
\affiliation{Department of Mechanical Engineering, Baylor University, Waco, TX {76798}, USA }




\date{\today}

\begin{abstract}
The oscillation of droplets supported by solid surfaces is important to a wide variety of applications such as dropwise condensation. 
In the present study, the axisymmetric natural oscillations of a liquid drop supported by a flat surface is investigated by direct numerical simulation. The liquid-gas interface is captured using a geometric volume-of-fluid (VOF) method. A parametric study is carried out by varying the equilibrium contact angle and the gravitational Bond number (Bo). Both positive and negative gravities are considered, and thus the results cover both pendant and sessile drops. To incorporate the effect of contact line mobility, the two asymptotic limits, namely the pinned contact line (PCL) and free contact line (FCL) conditions, are considered and their effects on the drop oscillation  features are characterized. The predicted oscillation frequencies for PCL and FCL serve as the upper and lower bounds for general situations. The drop oscillation is initiated by increasing the gravity magnitude for a short time. The first mode due to the drop centroid translation dominates the excited oscillation. The oscillation frequency scales with the capillary frequency, and the normalized frequency monotonically decreases with the equilibrium contact angle. For zero gravity, the computed frequencies for all contact angles agree remarkably well with the inviscid theory for both the PCL and FCL conditions. The kinetic energy correction factor is introduced to account for the additional contribution of the oscillation-induced internal flow to the overall kinetic energy of the drop. Both the frequency and the kinetic energy correction factor increase with Bo, decrease with the contact angle, and increase when the contact line condition changed from FCL to PCL. The variation of oscillation frequency due to the change of Bo is particularly significant when the contact angle is large, suggesting that the gravity effect must be incorporated to accurately predict the oscillation frequency for drops supported by hydrophobic or superhydrophobic surfaces. 
\end{abstract}

\maketitle
\section{Introduction}
\label{sec:intro}

The oscillation of droplets supported by solid surfaces is important to a wide variety of applications, such as drop shedding on condensation surfaces \cite{Yao_2017a} and water harvesting \cite{Dai_2018a}. When the surface normal vector is aligned and opposite of the gravity direction, the supported drops are also referred to as pendant and sessile drops, respectively. Drop oscillation induced by mechanical surface vibrations or acoustics has been shown to enhance drop mobility on the supported surface \cite{Yao_2017a}, or even to cause drops to detach from the surface \cite{Boreyko_2009a}. Due to the resonance effect, when the frequencies for the external forcing match the natural frequencies of the supported drops, the excited oscillation amplitude will be maximized for a given energy input \cite{Boreyko_2009a, Yao_2017a}. Therefore, it is advantageous to accurately predict the natural oscillation frequencies for supported drops on surfaces of different material properties. 

The natural oscillation of a liquid drop, when there is no external forcing, is a classic fluid mechanics problem and has been extensively studied in the past. For a free isolated drop, Rayleigh provided the explicit expression for the oscillation frequency for a given mode $n$ in the inviscid, free-surface, and small-amplitude limit \cite{Rayleigh_1879a}. The Rayleigh frequency $\omega_\text{Ra}$ scales with the capillary frequency $\omega_c$, and the ratio $\omega/\omega_c$ is a function of the mode number $n$, 
\begin{equation}
	\frac{\omega_\text{Ra}^2}{\omega_{c}^2} = (n-1)n(n+2)
	\label{eq:Rayleigh_freq}
\end{equation}
for $n\ge 2$. The capillary frequency is defined as $\omega_{c}=\sqrt{\sigma/(\rho_l R_0^3)}$,
 where $\sigma$ and $\rho_l$ are the surface tension and the liquid density, {respectively}, and $R_0$ is the radius of the spherical drop. The effect of the surrounding fluid has been incorporated by Lamb \cite{Lamb_1932a} and the Lamb frequencies can be expressed as 
\begin{equation}
	\frac{\omega_\text{Lamb}^2}{\omega_{c}^2} = \frac{(n-1)n(n+1)(n+2)}{(n+1)+n \rho_g/\rho_l}\,,
\end{equation}
where $\rho_g$ is the density of the surrounding gas. For cases with a small density ratio $\rho_g/\rho_l\ll 1$, the difference between the Lamb and Rayleigh frequencies is small. 
 
The effect of liquid viscosity on drop oscillation is generally characterized by the Ohnesorge number ($\text{Oh}$). For drops with finite $\text{Oh}$, the oscillation amplitude will decrease over time due to viscous dissipation. For small-amplitude oscillations, the decay of oscillation amplitude follows an exponential function $A(t) \sim \exp(-\beta t)$, where $\beta$ is the damping rate, which scales with the viscous frequency $\omega_v=\nu_l/R_0^2$, where $\nu_l$ is the kinematic viscosity of the drop liquid. The normalized damping rate $\beta/\omega_v$ is also a function of the mode number as shown by Lamb \cite{Lamb_1932a},
\begin{equation}
	\frac{\beta_\text{Lamb}}{\omega_v} = (n-1)(2n+1)\,. 
	\label{eq:Beta_Lamb}	
\end{equation}
The oscillation frequency generally decreases with $\text{Oh}$, though quite slowly. The leading order correction to the oscillation frequency is quadratic, $\omega^2=\omega_\text{Ra}^2-\beta^2$, which can be expanded as
\begin{equation}
	\frac{\omega^2}{\omega_{c}^2} = (n-1)n(n+2) -  (n-1)^2(2n+1)^2\ \text{Oh}^2 + O(\text{Oh}^3)\,,
\end{equation}
It is observed that, for drops with low $\text{Oh}$, the viscous effect on oscillation frequency is small unless the mode number $n$ is very large. 

The aforementioned studies all assume the drop oscillation amplitude is small. As a result, the oscillation is linear and a  superposition of different oscillation modes is allowed. When the oscillation amplitude is finite, the oscillation becomes nonlinear \cite{Trinh_1982a,Tsamopoulos_1983a} and the additional effects such as inter-mode coupling arise \cite{Basaran_1992a}. Furthermore, when a drop is moving, such as falling under the action of gravity, the surrounding gas flows can also influence the drop oscillation \cite{Helenbrook_2002a,Bergeles_2018a,Zhang_2019a}. In the present study, the focus is on low Ohnesorge-number drops (water droplets of millimeter sizes) and small-amplitude oscillations (oscillation amplitude lower than 10\% of drop radius). Furthermore, the drop liquid density and viscosity are significantly larger than those of the surrounding gas. As a result, if the drop is not supported by the solid surface, the oscillation frequency and the damping rate are expected to follow the Rayleigh frequencies and Lamb's damping rate. 

When the drop is in contact {with} and supported by a solid surface, additional complexities arise due to the interaction between the drop and the surface. First of all, the supported drop exhibits a first-mode ($n=1$) oscillation, which is associated with the drop centroid translation \cite{Strani_1984a,Sakakeeny_2020a}. For a free drop, the first mode and the corresponding centroid motion does not trigger a shape deformation, if the effect of ambient fluid is ignored. For a supported drop, however, when the distance between the drop centroid and the surface varies, there must be a corresponding deformation of the drop surface \cite{Sakakeeny_2020a}. 

\begin{figure}
 \centering
 \includegraphics[width=0.8\textwidth]{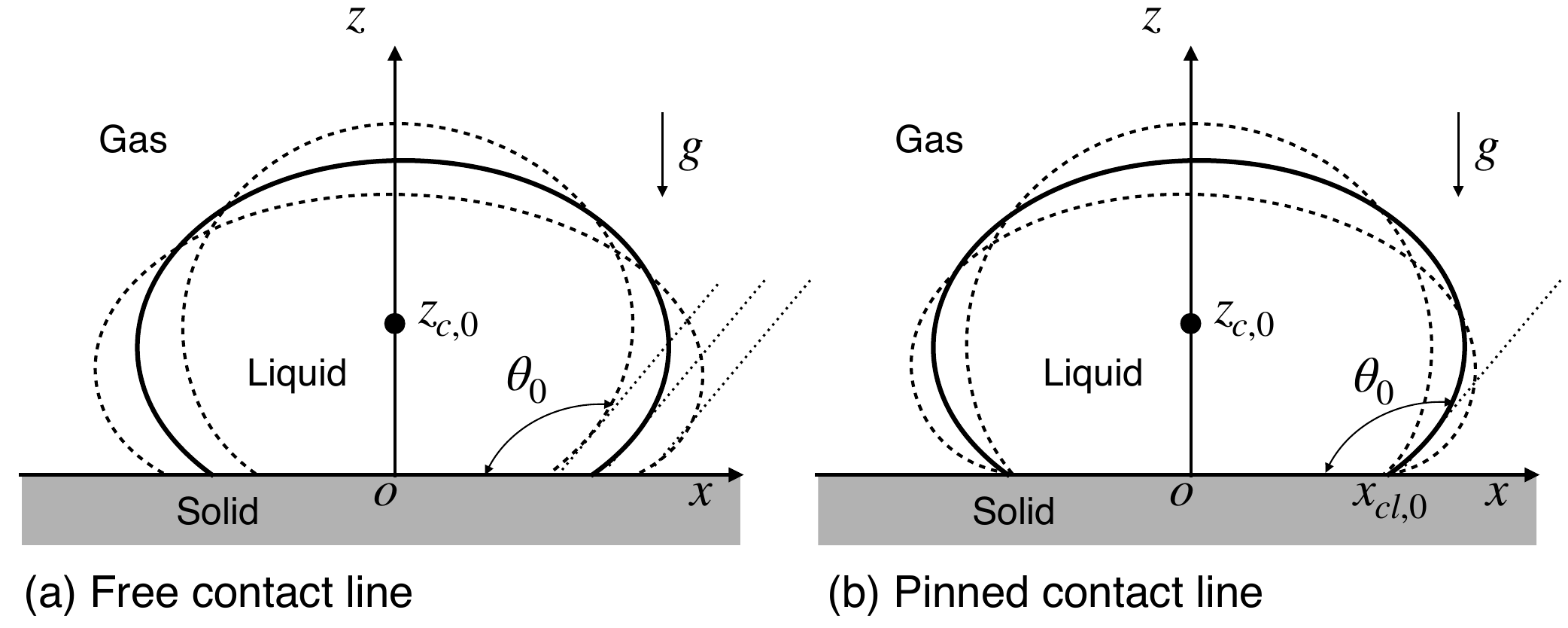}
 \caption{Schematics of supported drop oscillations with (a) free contact line (FCL) and (b) pinned contact line (PCL), where the solid lines represent the equilibrium state. }
 \label{fig:Schem}
\end{figure}

Furthermore, the natural oscillations of the supported drop will also be influenced by the surface material properties, such as the equilibrium contact angle and contact-line dynamics \cite{Strani_1984a, Noblin_2004a, Chang_2013b, Bostwick_2014a}. 
{Oscillation of supported drops can induce motion of the contact line. When contact-line hysteresis is present, the contact angle varies as the contact line moves. The angle for an advancing contact line is typically larger than that for the receding counterpart. The effect of hysteresis is typically characterized by the difference between the advancing and receding contact angles.  Modeling moving contact lines in continuum mechanics remains an unresolved challenge \cite{Bonn_2009a,He_2019a}. The present study is focused only on the two asymptotic limits for the contact-line mobility: the pinned contact line (PCL) and the free contact line (FCL), see Figs.\ \ref{fig:Schem} (a) and (b), respectively, where the typical singularity behaviors for moving contact lines  \cite{Snoeijer_2013a} are alleviated. 
For PCL, it is considered that the hysteresis effect is strong and the oscillation amplitude is small, so the contact angle always lies between the receding and advancing contact angles. As a result, the contact line is fixed/pinned while the contact angle can vary. For FCL, it is considered that there is no hysteresis effect, so the contact line can move freely while the contact angle is fixed at its equilibrium value. The oscillation frequencies for supported drops with general contact-line conditions will be bounded by these two limits.}
The previous inviscid theoretical models \cite{Strani_1984a,Bostwick_2014a,Sakakeeny_2020a} all indicated that the oscillation frequencies decrease with the contact angle for all oscillation modes. Bostwick and Steen \cite{Bostwick_2014a} further indicated that the oscillation frequency for a given mode number and contact angle increases significantly if the contact-line condition changes from FCL to PCL. 

Finally, the gravity also affects the oscillation of a supported drop. In the present study, it is considered that the surface is flat and the gravity is normal to the surface. It is taken that the gravity $g$ is positive when it is opposite to the surface normal. Therefore, positive and negative values of $g$ represent sessile drops and pendant drops, respectively. The effect of gravity can be characterized by the gravitational Bond number $\text{Bo} = \rho_l g R_d^2 / \sigma$, where {$R_d=  (3 V_d / 4 \pi)^{1/3}$ is the volume-based radius and $V_d$ is the volume of the drop}. The Bond number serves as a measure for the ratio between the gravity and surface tension contributions. Previous studies of supported drop oscillations often ignore the effect of $\text{Bo}$  \cite{Strani_1984a, Bostwick_2014a}. Nevertheless, numerical studies have shown that the oscillation frequency increases with $\text{Bo}$ for sessile drops \cite{Sakakeeny_2020a} and decreases with the magnitude of $\text{Bo}$ for pendant drops \cite{Basaran_1994a}. With the present definitions for $g$, Bo varies from negative to positive values, and the normalized oscillation frequency will then increase monotonically with $\text{Bo}$, though a more comprehensive investigation is required. 

The goal of the present study is to investigate the natural oscillation of a liquid drop supported by a flat solid surface through {direct} numerical simulation. {The oscillation is excited by increase the gravity magnitude for a short time. Since the normal of the surface is taken to be aligned with the gravity, the induced oscillation is axisymmetric.} Particular attention is {paid to} the first oscillation mode, because it generally dominates the excited oscillations. The effects of contact angle, contact line mobility, and gravitational Bond number on important oscillation features, including the oscillation frequency, the damping rates, and the oscillation-induced internal flow, will be characterized through parametric simulations. As an extension {of} our former study on oscillation of sessile drops with FCL condition \cite{Sakakeeny_2020a}, the present study is focused on the PCL condition, so that a comprehensive understanding of the effect of contact-line mobility on the oscillation of supported drops can be established. We will vary the Bond number for both negative and positive values, to cover both the pendant and sessile drops regimes. 

The rest of the paper will be organized as follows. The simulation approaches, including the governing equations, the numerical methods, and the simulation setup, will be presented in Section \ref{sec:sim}. The simulation results will be shown and discussed in Section \ref{sec:results}. Finally, we will conclude the key findings in Section \ref{sec:conclusions}.

\section{Simulation methods}
\label{sec:sim}
\subsection{Governing equations}
The liquid-gas two-phase flow is resolved using the one-fluid approach, wherein the two phases, liquid and gas, are treated as one fluid with material properties that change abruptly across the gas-liquid interface. The Navier-Stokes equations for incompressible flow with surface tension are given as 

\begin{equation}
  \rho (\partial_t \mathbf{u} + \mathbf{u} \cdot \nabla \mathbf{u}) = -\nabla p + \nabla \cdot (2 \mu \mathbf{D}) + \sigma \kappa \delta_s \mathbf{n},
  \label{eq:NS1}
\end{equation}
\begin{equation}
  \nabla \cdot \mathbf{u} = 0,
  \label{eq:NS2}
\end{equation}
 where $\rho$, $\mathbf{u}$, $p$, and $\mu$, represent density, velocity, pressure, and viscosity, respectively. The strain-rate tensor is denoted by $\mathbf{D}$. The surface tension term on the right-hand side of Eq.\ \eqr{NS1} is a singular term, with the Dirac distribution function $\delta_s$ localized on the interface. The surface tension coefficient is represented by $\sigma$, and $\kappa$ and $\mathbf{n}$ are the local curvature and unit normal of the interface, respectively.
 
The two different phases are distinguished by the liquid volume fraction $C$. While $C=0$ indicates that the cell is full of gas, $C=1$ indicates that the cell is full of liquid. For cells with interfaces, $0 < C < 1$. The temporal evolution of $C$ satisfies the advection equation, 
\begin{equation}
  \partial_t C + \mathbf{u} \cdot \nabla C = 0.
  \label{eq:C1}
\end{equation}

The fluid density and viscosity are determined by 
\begin{align}
  \rho & = C \rho_l + (1 - C) \rho_g\,,
  \label{eq:density1}\\
  \mu &= C \mu_l + (1 - C) \mu_g\, ,
  \label{eq:viscosity1}
\end{align}
where the subscripts $g$ and $l$ correspond to the gas and the liquid phases, respectively.

\subsection{Numerical methods}
The governing equations (Eqs.\ \eqr{NS1}, \eqr{NS2}, and \eqr{C1}) are solved using the open-source, multiphase flow solver \textit{Basilisk} \cite{basilisk}. The \textit{Basilisk} solver uses a finite-volume approach based on a projection method. An adaptive quadtree spatial discretization is used, which allows for adaptive mesh refinement (AMR) in user-defined regions. The advection equation (Eq.\ \eqr{C1}) is solved via the piecewise-linear geometrical {volume-of-fluid} (VOF) method \cite{Scardovelli_1999a, Popinet_2009a}. 
{Compared to other popular interface-capturing methods, such as the front-tracking \cite{Unverdi_1992a}, level-set \cite{Sussman_1994a}, the VOF method has the important advantage of conserving mass/volume, which is crucial to predicting oscillation frequency, as desired in the present study, since the frequency is a function of the drop volume. The surface tension calculation in VOF framework can induce numerical parasitic currents near the interface \cite{Renardy_2002a}. This numerical issue is solved by combining the balanced-force continuum-surface-force method for surface tension discretization and the height-function (HF) method for curvature calculation \cite{Popinet_2009a}.} The HF method is additionally used to specify the contact angle at the surface. The \textit{Basilisk} solver utilizes a staggered-in-time discretization of the volume fraction/density and pressure, leading to a formally second-order-accurate time discretization \cite{Popinet_2009a}. Numerous validation studies for the numerical methods, as well as examples of a wide variety of interfacial multiphase flows, can be found on the \textit{Basilisk} website and in previous studies \eg, \cite{Zhang_2019b, Marcotte_2019a, Zhang_2020a, Mostert_2020a, Sakakeeny_2020a}.

\subsection{Physical parameters}
\label{sec:parameters}

\begin{figure}
 \centering
 \includegraphics[width=0.8\textwidth]{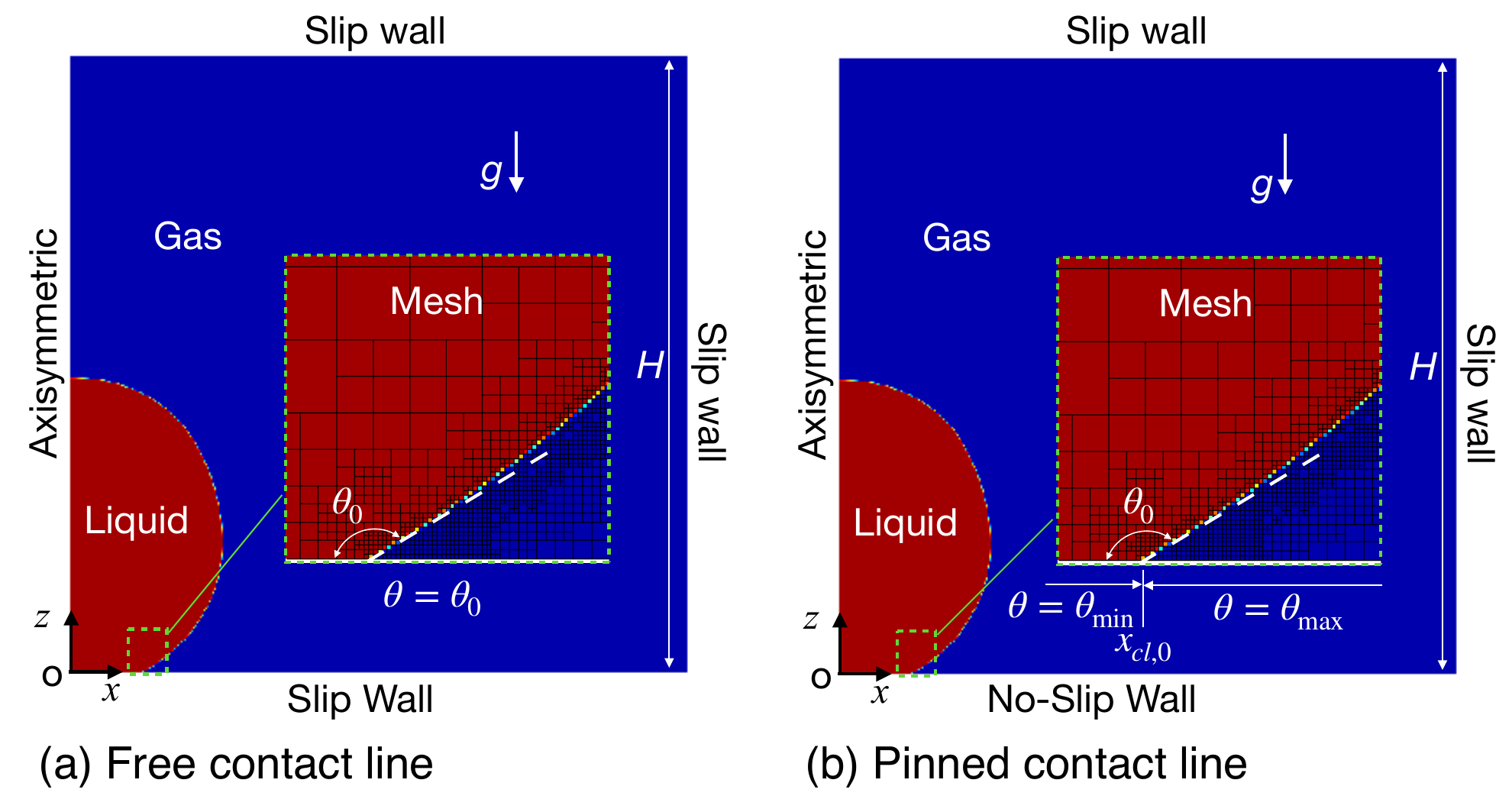}
 \caption{Simulation setup for (a) free contact line (FCL) and (b) pinned contact line (PCL) cases.}
 \label{fig:SimSetup}
\end{figure}

In the present study, we consider the axisymmetric natural oscillations of a viscous liquid drop supported by a flat surface, as shown in Fig.\ \ref{fig:SimSetup}. The physical properties for the liquid and gas phases are taken to be similar to water and air, respectively. The volume of the drop is kept constant, $V_d=65.45$ \textmu L, across all cases, for which the volume-based radius $R_d=2.5$ mm. The wettability of the surface is characterized by the equilibrium contact angle, $\theta_0$, which is varied from 50 to $\ang{150}$. The range of contact angles considered here is sufficient to cover common hydrophilic, hydrophobic, and superhydrophobic surfaces \cite{Yao_2017a}. 
The values of the key physical parameters are listed in Table \ref{tab:physParam}. 

{While the equilibrium shape for a free drop is a sphere, the equilibrium shape for a supported drop is a spherical cap, when gravity is absent. The the radius of the spherical cap is also denoted by $R_0$, which varies with $\theta_0$ for a given volume $V_d$ as
\begin{align}
    R_0 & = \left(\frac{3V_d}{\pi (2 + cos \theta_0) (1-cos \theta_0)^2}\right)^{1/3} \,. 
    \label{eq:R0}
\end{align}
Since $R_0$ better represents the surface curvature, the capillary frequency is defined based on $R_0$ as $\omega_c = \sqrt{\sigma / (\rho_l R_0^3)}$.
}

The key dimensionless parameters, defined based on scaling variables $R_d$, $\rho_l$, and $\sigma$, are listed in Table \ref{tab:physParamNonDim}. {It} can be seen that the gas-to-liquid ratios for density and viscosity are quite small, thus the effect of the surrounding gas on the liquid drop is minimal. The Ohnesorge number $\text{Oh}=0.0024$ indicates that the effect of viscosity is weak. 
Furthermore, variation of $\text{Oh}$ due to moderate change of drop volume will have little effect on the normalized oscillation frequency, $\omega/\omega_c$, where $\omega$ and $\omega_c$ are the drop oscillation and capillary frequencies, respectively. {For this reason,} we have considered only one drop volume. 

The effect of gravity is characterized by the Bond number, $\text{Bo}$. The value of $\text{Bo}$ can be varied by changing $V_d$ or $g$. In the present study, we keep $V_d$ fixed and vary $g$ from -0.98 to 9.8 m/s$^2$. Negative $\text{Bo}$ and $g$ represent the cases for pendant drops. The resulting range of $\text{Bo}$ is -0.088 to 0.88. {For pendant drops with large $\theta_0$, the drop can be unstable and detach from the surface if $|\text{Bo}|$ is large. Therefore, a smaller range of Bo is considered for the pendant drop than the sessile drop. It is confirmed that for the range of Bo considered, the equilibrium state of the supported drop is stable for all $\theta_0$ considered. We have also considered only small-amplitude oscillations, so that the drop will not detach from the wall.} 

While Oh and Bo are defined based on $R_d$, the Bond and Ohnesorge numbers can be alternatively defined based on $R_0$ as $\text{Bo}_0=\rho_l g R_0^2 / \sigma$ and $\text{Oh}_0=\mu_l / (\rho_l \sigma R_0)$, which will then vary with $\theta_0$. 

\begin{table}[tbp]
    \centering
    \begin{tabular}{c c c c c c c c}
        \hline
        \begin{tabular}{@{}c@{}}$\rho_l$ \\ $(kg/m^3)$ \\ \end{tabular} & 
        \begin{tabular}{@{}c@{}}$\rho_g$ \\ $(kg/m^3)$ \\ \end{tabular} & 
        \begin{tabular}{@{}c@{}}$\mu_l$  \\ $(Pa \cdot s)$ \\ \end{tabular} &
        \begin{tabular}{@{}c@{}}$\mu_g$  \\ $(Pa \cdot s)$ \\ \end{tabular} & 
        \begin{tabular}{@{}c@{}}$\sigma$ \\ $(N/m)$ \\ \end{tabular} &
        \begin{tabular}{@{}c@{}}$V_{d}$ \\ $(\mu L)$ \\ \end{tabular} & 
        \begin{tabular}{@{}c@{}}$\theta_0$ \\ (\textdegree) \\ \end{tabular} &
        \begin{tabular}{@{}c@{}}$g$ \\ (m/s$^2$) \\ \end{tabular} \\
        \hline
         $1000$ & $1.2$ & $1 \times 10^{-3}$ & $1 \times 10^{-5}$ & $0.07$ & $65.45$  & 50 to 150 & -0.98 to 9.8\\
         \hline
    \end{tabular}
    \caption{Physical parameters.}
\label{tab:physParam}
\end{table}

\begin{table*}[tbp]
    \centering
    \begin{tabular}{c c c c c c}
        \hline
        r & $m$ & $\text{Oh}$ & $\theta_0$ & $\text{Bo}$\\
        \hline
        $\rho_g / \rho_l$ & $\mu_g / \mu_l$ & $\mu_l / \sqrt{\rho_l \sigma R_d}$ &(\textdegree) & $\rho_l g R_d^2 / \sigma$  \\
        \hline
         $0.0012$ & {$0.01$} & 0.0024 & {50} to {150} & -0.088 to 0.88\\
         \hline
    \end{tabular}
    \caption{Key dimensionless parameters.}
\label{tab:physParamNonDim}
\end{table*}

\subsection{Simulation setup}
\subsubsection{Computational domain and boundary conditions}
The computational domain is the same for all cases, see in Fig. \ref{fig:SimSetup}. The length of the square domain edge is $H=4R_d$. The axisymmetric boundary condition is applied on the left surface, while the top and right surfaces are slip walls. The drop is in contact with the bottom surface. For the FCL condition, the contact angle {on the bottom surface} is fixed as the equilibrium contact angle, namely $\theta=\theta_0$, see Fig. \ref{fig:SimSetup}(a), and the contact line can move freely when the drop oscillates. The interface normal for a given contact angle is specified using the height function method \cite{Afkhami_2009a}. To {alleviate} the singular behavior at the contact line, such as the diverging viscous stress \cite{Snoeijer_2013a}, the bottom surface for FCL cases is taken to be a slip wall. 

{\subsubsection{Pinned-contact-line boundary conditions}}
For the PCL condition, the contact line is pinned at its equilibrium position $x_{cl,0}$ for a given Bo, and the contact angle can vary freely when the drop oscillates. To be consistent with the contact line condition, we treat the bottom surface as a no-slip wall, see Fig.~\ref{fig:SimSetup}(b). {To pin the contact line at a given location $x_{cl,0}$, the contact angle on the bottom surface is specified as $\theta=\theta_{\min}$ for $x<x_{cl,0}$ and $\theta=\theta_{\max}$ $x>x_{cl,0}$, see Fig.\ \ref{fig:SimSetup}(b).  The contact angle needs to reach $\theta_{\min}$ and $\theta_{\max}$ for the contact line to move to the the left and right, respectively. If $\theta$ varies between $\theta_{\min}$ and $\theta_{\max}$ when the drop oscillates, then the contact line will not move. For the present study, the equilibrium contact angle $\theta_0$ is varied from $\ang{50}$ to $\ang{150}$. For small-amplitude oscillations, the contact angle $\theta$ only varies  in a small extent around $\theta_0$. As long as $\theta_{\min}$ and $\theta_{\max}$ are sufficiently small and large, the specific values are immaterial. In the present study, we have set $\theta_{\min}=\ang{15}$ and $\theta_{\max}=\ang{165}$, which are shown to be sufficient to statisfy $\ang{15} \ll \theta \ll \ang{165}$ for all cases considered and to pin the contact line effectively when the drop oscillates. Though the abrupt change of contact angle on the bottom surface can effectively pin the contact line, the boundary condition introduces small-amplitude velocity fluctuations in the cell where the contact line is located, which leads to non-physical kinetic energy. To eliminate this numerical artifact, the fluid velocity in the cells that are $l_{0}$ away from the contact line location is manually set to be zero in every time step.  For all the simulations, $l_0/\Delta_{\min}=4$, which have been verified to effectively eliminate the numerical velocity oscillation without influencing the oscillation dynamics.}

\subsubsection{Initial conditions}
For a given combination of $\theta_0$ and $\text{Bo}$, the initial shape of the supported drop is taken to be the equilibrium shape. The geometry of the equilibrium supported drop and the contact line location $x_{cl,0}$ can be obtained by the equilibrium drop theory and numerically solving a system of ODE. The details can be found in our previous study and thus are not repeated here \cite{Sakakeeny_2020a}. To initiate the shape oscillation, the gravity magnitude is increased for a short duration. The magnitude of gravity $|g|$ is increased by $g_{pert}$ for $t\le t_{pert}$. The perturbation gravity $g_{pert}=4.9$ m/s$^2$ and the perturbation time $t_{pert}=0.14 \sqrt{\rho_l R_d^3/\sigma}$ for all cases. Due to the change of gravity, the drop will be pushed down ($g>0$) or pulled up ($g<0$) and deviate from the equilibrium shape. Once the gravity returns to the original value for $t>t_{pert}$, the drop deforms toward the equilibrium shape and starts to oscillate. {Since the surface normal is aligned with the gravity, only the axisymmetric zonal modes will be excited.} Though all oscillation modes will be excited to some extent by this method, the first mode ($n=1$) dominates other high-order modes. 

\subsubsection{Mesh resolution}
A quadtree adaptive mesh is used to discretize the domain. The local cell size is adapted based on the estimated discretization errors of the liquid volume fraction and the velocity components. The assessment of discretization error for each variable is made through a wavelet transform \cite{Hooft_2018a}. If the estimated error is larger than the specified threshold, the mesh will be locally refined, or vice versa. For the present simulation, the normalized error thresholds for the volume fraction and the velocity are set as 0.001 and 0.0001, respectively. Tests have been made to verify these thresholds are sufficiently small. A representative snapshot of the  mesh close to to the contact line is shown in Fig. \ref{fig:SimSetup}(c). The minimum cell size in the quadtree mesh is controlled by the maximum refinement level, $L$, i.e., $\Delta x_{\min}=H/2^L$. The mesh for $L=11$ is used in the present simulation, which corresponds to $R_0/\Delta x_{\min} \approx 512$, namely 512 cells across the drop radius.

\subsection{Summary of simulation cases}
To systematically investigate the effects of the equilibrium contact angle ($\theta_0$), the Bond number ($\text{Bo}$), and the contact line mobility on the oscillation of a sessile drop, 11 different values of $\theta_0$ (from $\ang{50}$ to $\ang{150}$ with an increment of $\ang{10}$) and 9 different values of $\text{Bo}$ (from -0.088 to 0.88) have been used for both the FCL and PCL conditions. Therefore, a total of 198 cases are simulated in the parametric study. 

The simulations were performed on the Baylor University cluster \emph{Kodiak} using 4 to 18  CPU cores (Intel E5-2695 V4). Each simulation case takes about 133 to 195 hours of CPU time to reach the time $t\omega_c\approx 105$ (51 s). The simulation time has been verified to be sufficiently long to measure the frequencies for the first oscillation mode.

\section{Results}
\label{sec:results}

\subsection{Grid refinement and validation}
\label{sec:validation_pcl}

\begin{figure}
 \centering
 \includegraphics[width=1.0\textwidth]{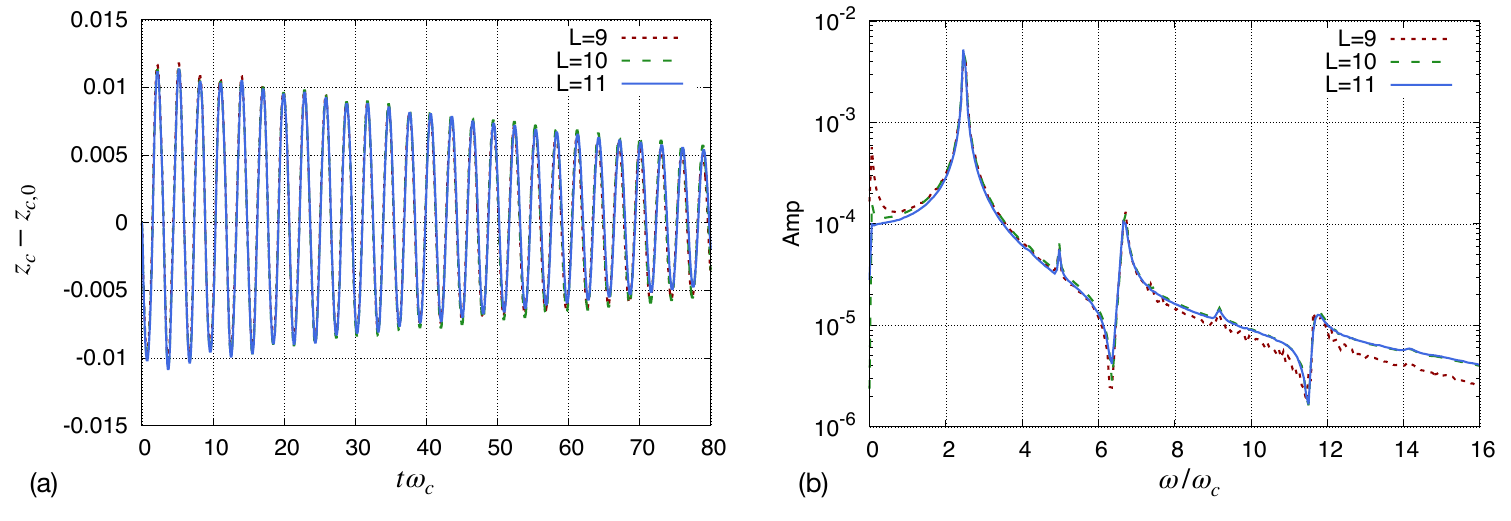}
 \caption{(a) Temporal evolution and (b) frequency spectrum of the drop centroid height $z_c$ for $\theta_0=\ang{130}$ and $\text{Bo}=0$, for three different mesh refinement levels $L=$9, 10, and 11.}
 \label{fig:ca130Bo0MeshRes_xc}
\end{figure}

A grid refinement study varying $L=9$ to 12 has been performed for $\theta_0=\ang{130}$ and $\text{Bo}=0$. The results for the temporal evolution of the drop centroid height $z_c$ and the corresponding frequency spectra obtained by the Fourier transform of the temporal signals are shown in Fig.\ \ref{fig:ca130Bo0MeshRes_xc}. The difference between the results for $L=11$ and 10 are almost invisible, demonstrating that the refinement level $L=11$ is sufficient to fully resolve the oscillations of supported drops. 

\begin{figure}%
    \centering
    \includegraphics[width=0.8\textwidth]{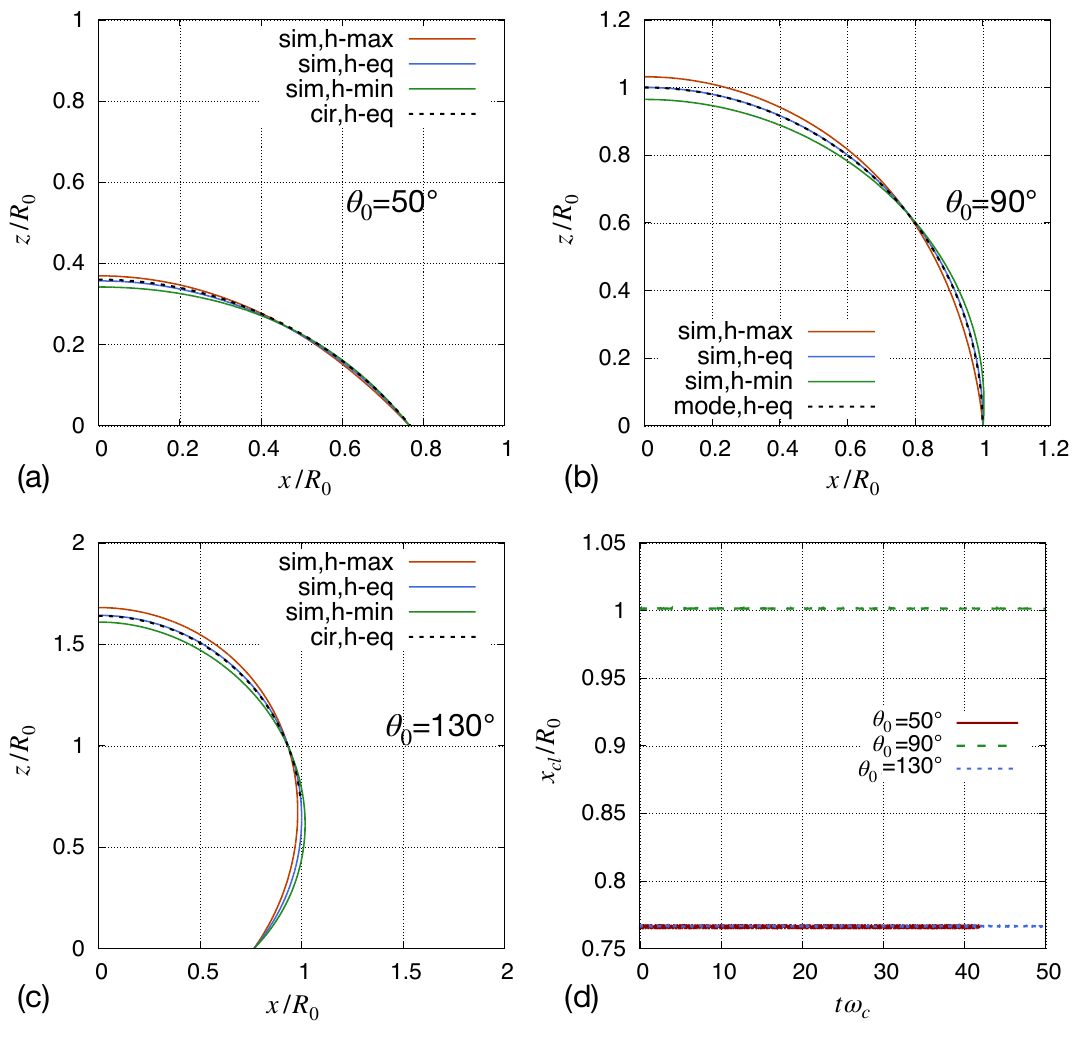}
    \caption{Validation study for the PCL condition. (a)-(c) Representative drop surfaces corresponding to the maximum, equilibrium, and minimum drop heights in one first-mode oscillation cycle for $\theta_0=\ang{50}$, $\ang{90}$, and $\ang{130}$. (d) Temporal evolution of the contact line location.}
    \label{fig:Validation_PCL}%
\end{figure}

Validation for the present simulation setup for the FCL condition can be found in our previous study \cite{Sakakeeny_2020a}. For additional validation of the PCL condition, we examine whether the contact line is effectively pinned when the drop oscillates. Representative drop surfaces corresponding to the maximum, equilibrium, and minimum centroid heights in one first-mode oscillation cycle for $\theta_0=\ang{50},\ang{90},\ang{130}$ are shown in Figs.~\ref{fig:Validation_PCL}(a)-(c), respectively. The temporal evolution of the contact line $x$-location $x_{cl}$ for $\theta_0=\ang{50},\ang{90},\ang{130}$ is shown in Fig.~\ref{fig:Validation_PCL}(d), and it is clearly shown that the contact line is successfully pinned for all cases shown. The equilibrium contact line locations for $\theta_0=\ang{50}$ and $\ang{130}$ are the same. 

\subsection{Oscillation frequency}
\label{sec:dropcentroidandfreqspectrum}
\begin{figure}
 \centering
 \includegraphics[trim=0 0 0 0, clip, width=\textwidth]{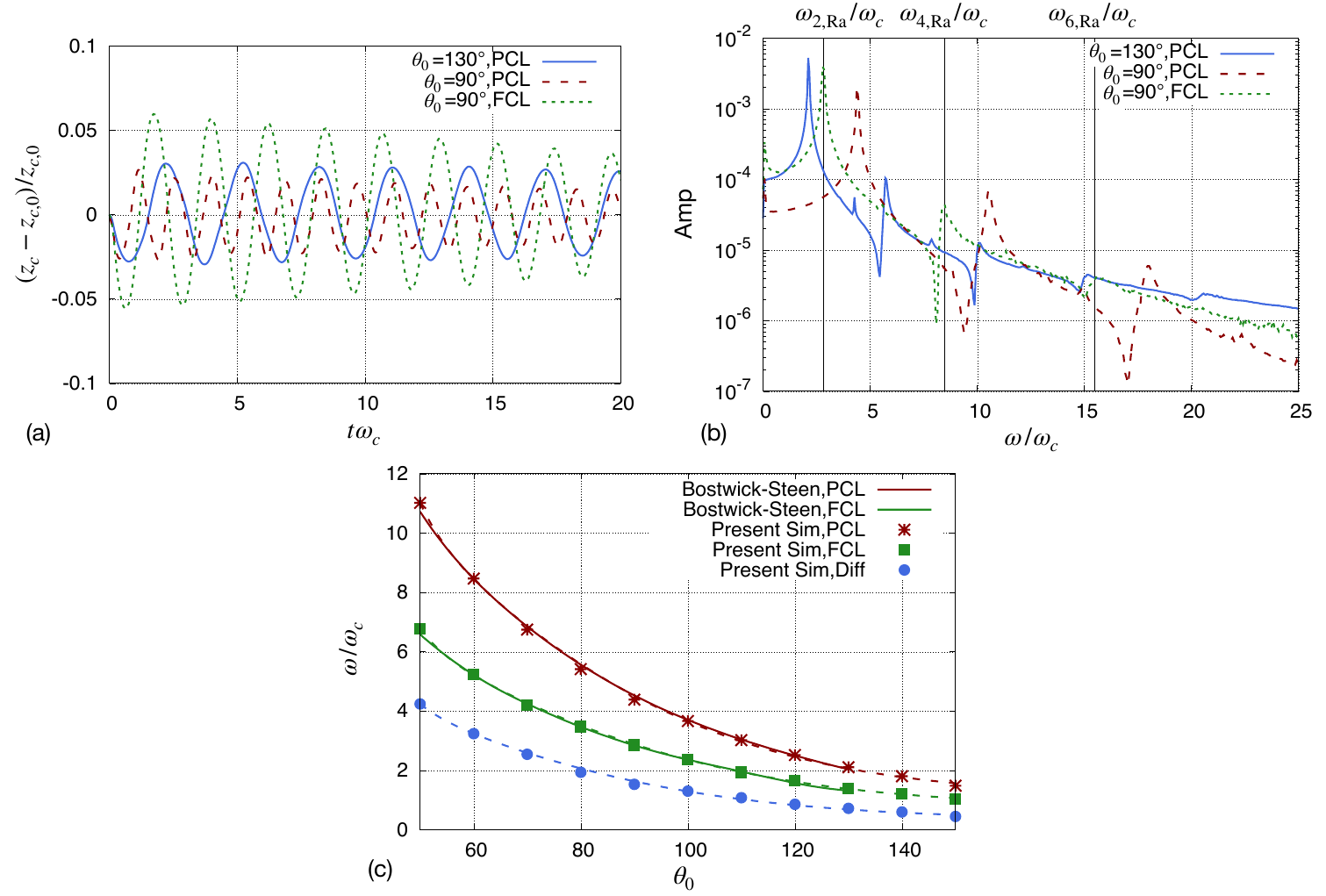}
 \caption{(a) Temporal evolution and (b) frequency spectrum of the drop centroid height for different $\theta_0$. (c) Variations of normalized first-mode oscillation frequency $\omega/\omega_c$ and the difference between the two $(\omega_P-\omega_F)/\omega_c$ as a function of contact angle $\theta_0$ for $\text{Bo}=0$. In (b) the Rayleigh frequencies (Eq.\ \eqr{Rayleigh_freq}) for the $n=2$, 4, and 6 modes are shown for comparison.  In (c) the simulation results (symbols) are compared with the inviscid theory of Bostwick and Steen \cite{Bostwick_2014a} (solid lines). The dashed lines represent fitting correlations for the simulation results.  }
 \label{fig:xc_IC1_Bo0_theta_combined}
\end{figure}

The frequency of the shape oscillation depends on the mode number. In the present study we focus on the dominant first mode $n=1$. The frequency of the $n=1$ mode can be measured through the temporal evolution of the centroid height $z_c$. The temporal evolution of $z_c$ for $\theta_0=\ang{90}$ and $\ang{130}$ and the PCL condition are shown in Fig.\ \ref{fig:xc_IC1_Bo0_theta_combined}(a). The drop for $\theta_0=\ang{90}$ and FCL is also shown for comparison. For all three cases shown here, $\text{Bo}=0$. It can be seen that $z_c$ oscillates with respect to the equilibrium value $z_{c,0}$. The oscillation amplitude is generally small compared to $z_{c,0}$ so the oscillation is expected to be linear. 

Fourier transform is performed to generate frequency spectra, which are used to identify oscillation frequencies (shown as peaks in the spectra). Higher-order modes $n>1$ are also observed in the spectra, though the first mode is clearly the dominant one. As addressed in previous studies \citep{Bostwick_2014a, Sakakeeny_2020a}, the flat surface for $\theta_0=\ang{90}$ and FCL is identical to the symmetric boundary condition. Therefore, a supported drop with $\theta_0=\ang{90}$ is equivalent to the top half of a free drop with twice the size. The oscillation frequency for the $n^{\text{th}}$ mode for a supported drop with $\theta_0=\ang{90}$ and FCL is identical to that for the $(2n)^{\text{th}}$ mode for a free drop. Since the oscillation frequencies for the free drop for the present fluid properties are well predicted by the Rayleigh frequencies, the values of $\omega_\text{Ra}$ for the $n=2$, 4, and 6 are plotted in Fig.\ \ref{fig:xc_IC1_Bo0_theta_combined}(b) for comparison. It is clearly shown that $\omega_1$, {$\omega_2$}, and $\omega_3$  for the supported drop with $\ang{90}$ and FCL agree very well with $\omega_{2,\text{Ra}}$, $\omega_{4,\text{Ra}}$, and $\omega_{6,\text{Ra}}$. 

The oscillation frequency depends on both the contact angle and the contact line mobility. It is shown that $\omega/\omega_c$ decreases from about 4.5 to 2.1 when $\theta_0$ increases from $\ang{90}$ to $\ang{130}$. For the same contact angle, $\theta_0=\ang{90}$, $\omega/\omega_c$ decreases from 4.5 to 2.8 when the contact line mobility changes from PCL to FCL. For the same initial shape and perturbation method,  the oscillation amplitude of $z_c$ for FCL is significantly larger than that for  PCL, since the constraint of the latter condition on the drop is stronger. Correspondingly, the amplitude for FCL in the spectrum is also higher than that for PCL. 

The first-mode frequency for a supported drop with both FCL and PCL for $\text{Bo}=0$ are shown as a function of the equilibrium contact angle $\theta_0$ in Fig.\ \ref{fig:xc_IC1_Bo0_theta_combined}(c). For both FCL and PCL conditions, $\omega/\omega_c$ monotonically decreases with $\theta_0$. The decrease is more profound for small $\theta_0$ (hydrophilic surfaces),  and is more gradual for large $\theta_0$ (hydrophobic or superhydrophobic surfaces). When  $\theta_0\to \ang{180}$, the supported drop approaches a free drop and the constraint from the surface disappears. As a result, the oscillation frequency will reduce to zero. When $\theta_0$ is close to $\ang{180}$, the decrease of $\omega$ over $\theta_0$ becomes very rapid \cite{Strani_1984a}. Yet, a detailed analysis of the asymptotic behavior of the frequency near the limit of $\theta_0\to \ang{180}$ is out of the scope of the present paper. 

For all $\theta_0$, $\omega/\omega_c$ for PCL is higher than that for FCL. The difference between the two, $(\omega_P-\omega_F)/\omega_c$, also decreases with $\theta_0$, where $\omega_P$ and $\omega_F$ represent the oscillation frequencies for the PCL and FCL, respectively. As $\theta_0$ increases, the contact area decreases. As a result, the constraint of the surface {on} the drop reduces, and the effect of contact line mobility conditions will also become less important. Since PCL and FCL represent the two asymptotic limiting conditions for the contact line mobility, the predicted frequencies for PCL and FCL shown here represent the upper and lower bounds for the first-{mode} oscillation frequencies for general situations. The results are useful to estimate the natural frequency of a supported drop on arbitrary material surfaces. 

Bostwick and Steen  \cite{Bostwick_2014a} have established an inviscid theoretical model to predict the oscillation frequency for supported drop at $\text{Bo}=0$. Since $\text{Oh}$ in the present case is small, the inviscid theory of Bostwick and Steen is expected to be a good approximation. Their theoretical predictions are available for $\ang{50} < \theta_0< \ang{130}$ and are plotted in Fig.\ \ref{fig:xc_IC1_Bo0_theta_combined}(c) for comparison. The agreement between the simulation and theoretical model is excellent for both the FCL and PCL conditions. The good agreement observed further validates the simulation results. 

For convenience of using the present results, correlations for first-mode oscillation frequencies for the PCL and FCL conditions as a function of $\theta_0$ are fitted in the following form: 
\begin{align}
	\log(\omega/\omega_c) = c_0+c_1 (1+\cos\theta_0) +\left[\exp\left(\frac{(1+\cos\theta_0) ^{c_2} }{c_3} \right)-1\right]\,.
	\label{eq:omega_fit}
\end{align}
The fitted constants are $[c_0, c_1, c_2, c_3]=[-0.0901,\, 1.15,\, 20.3,\, 2.16\times 10^5]$ for FCL and  $[0.287,\, 1.22,\, 20.7,\, 2.81\times 10^5]$ for PCL. The fitting correlations are plotted in Fig.~\ref{fig:xc_IC1_Bo0_theta_combined}(c) and are shown to well represent the simulation results. For large $\theta_0$, $1+\cos(\theta_0)$ is small, and the expression above reduces to a linear function, \eg, $\log (\omega/\omega_c) \approx c_0+c_1 (1+\cos\theta_0) $. The linear relation between $\log (\omega/\omega_c)$ holds for all hydrophobic cases $\theta>\ang{90}$. The correction term $[\exp\left({(1+\cos\theta_0) ^{c_2} }/{c_3} \right)-1]$ is mainly used to account for the deviation of the hydrophilic cases from the linear function. It is also worth {mentioning} that the correlation Eq.\ \eqr{omega_fit} is strictly valid for the range of $\theta_0$ studied. It is not intended to capture the asymptotic behavior at $\theta_0=\ang{180}$, \ie, {where} the frequency drops rapidly over $\theta_0$ near singularity location $\theta_0=\ang{180}$ \cite{Strani_1984a}.

\subsection{Kinetic energy correction factor}
\label{sec:EkandEkc}
The kinetic energy of the liquid drop can be expressed as 
\begin{equation}
	E_k = \rho_l\int_{V_d} \frac{ |\bs{u}|^2}{2} dV \, ,
 \label{eq:EK1}
\end{equation}
which will vary over time as the drop oscillates. The temporal variation of $E_k$ is due to two contributions. The first contribution is related to the bulk motion of the drop following the velocity of the drop centroid, 
\begin{equation} 
	E_{kc}= \frac{m_d |\bs{u_c}|^2}{2} =  \frac{m_d w_c^2}{2}\, .
\end{equation}
where $w_c$ is the $z$-component of centroid velocity, $\bs{u_c}$. For a free drop, $E_{k}=E_{kc}$ since the translation of the drop does not induce shape deformation. However, $E_k>E_{kc}$ for a supported drop due to the additional contribution of the internal flow induced by the shape oscillation. 

\begin{figure}%
    \centering
    \includegraphics[width=\textwidth]{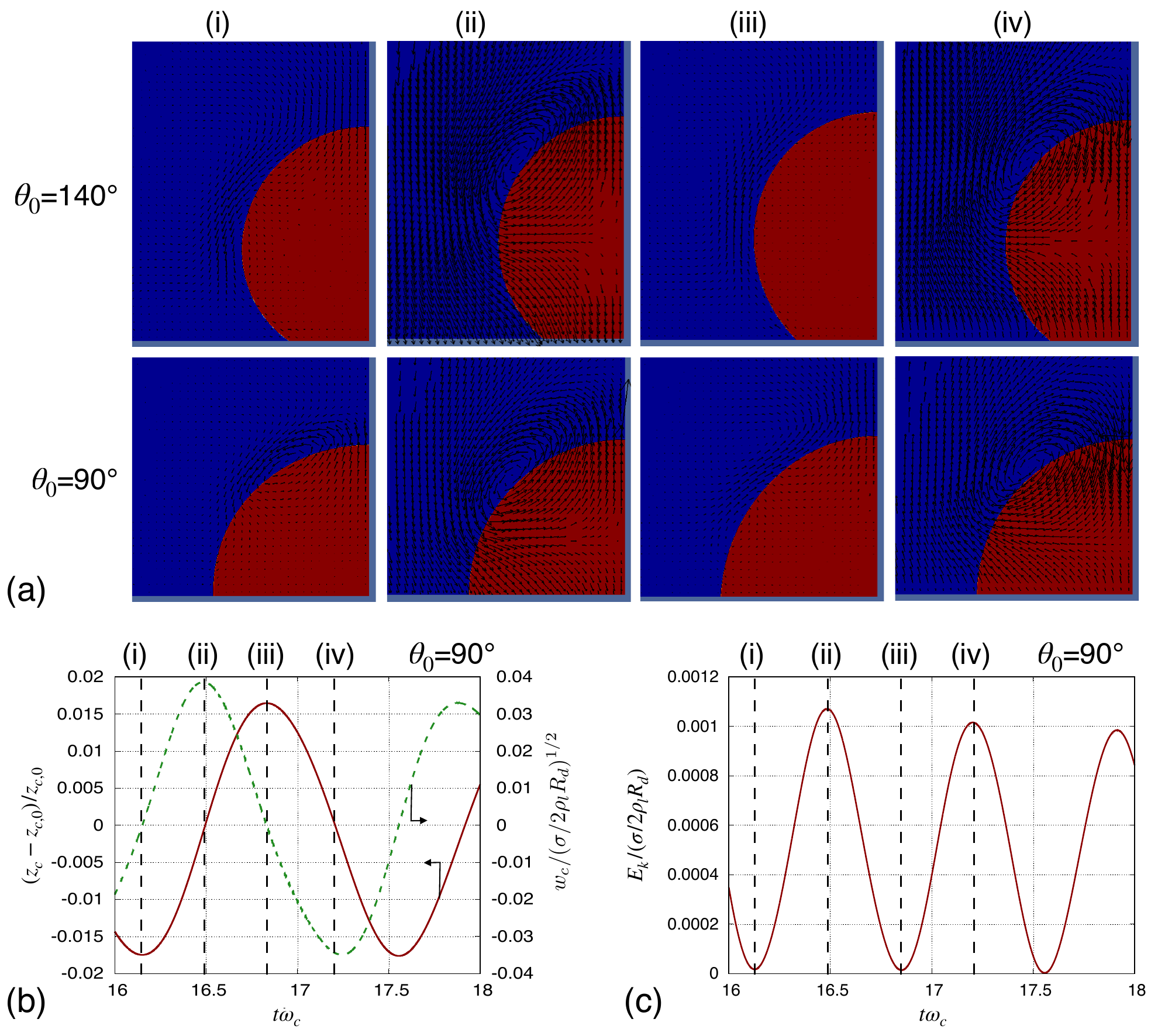}
    \caption{Temporal evolutions of (a) the flow field in the drop reference frame, (b) the centroid height $z_c$ and velocity $u_c$ and (c) the droplet kinetic energy $E_k$ for $\text{Bo}=0$ and PCL condition. The snapshots of the flow fields in (a) are for $\theta_0=\ang{90}$ and $\ang{140}$ and four critical phases of an oscillation cycle, as indicated in (b) and (c). }
    \label{fig:Flow_KE}%
\end{figure}

The velocity fields around the drop for different contact angles are shown in Fig.\ \ref{fig:Flow_KE}(a). The snapshots shown in Fig.\ \ref{fig:Flow_KE}(a) correspond to the valley, peak, and two equilibrium positions of the centroid in an oscillation cycle of the dominant $n=1$ mode, which are also indicated in the time evolutions of $z_c$ and $E_k$ in Figs.\ \ref{fig:Flow_KE}(b) and (c). The velocity here is in the drop reference frame and thus the contribution of the bulk motion has been subtracted. When the drop moves upward, see column (ii), the typical straining flow pattern can be recognized inside the drop. Furthermore, a clockwise circulation is generated on the left top corner of the drop. When the drop moves downward, see column (iv), the directions of the circulation and internal straining flow reverse. The flow pattern for $\theta=\ang{140}$ is quite similar to that for a free drop undergoing a $n=2$ mode oscillation. This similarity in the drop shape for supported drops with large $\theta_0$ has been observed by Strani and Sabetta \cite{Strani_1984a}. When $\theta_0$ decreases, such as $\theta=\ang{90}$, the flow pattern will {become} less similar to the free drop $n=2$ mode. 

\begin{figure}
 \centering
 \includegraphics[width=\linewidth]{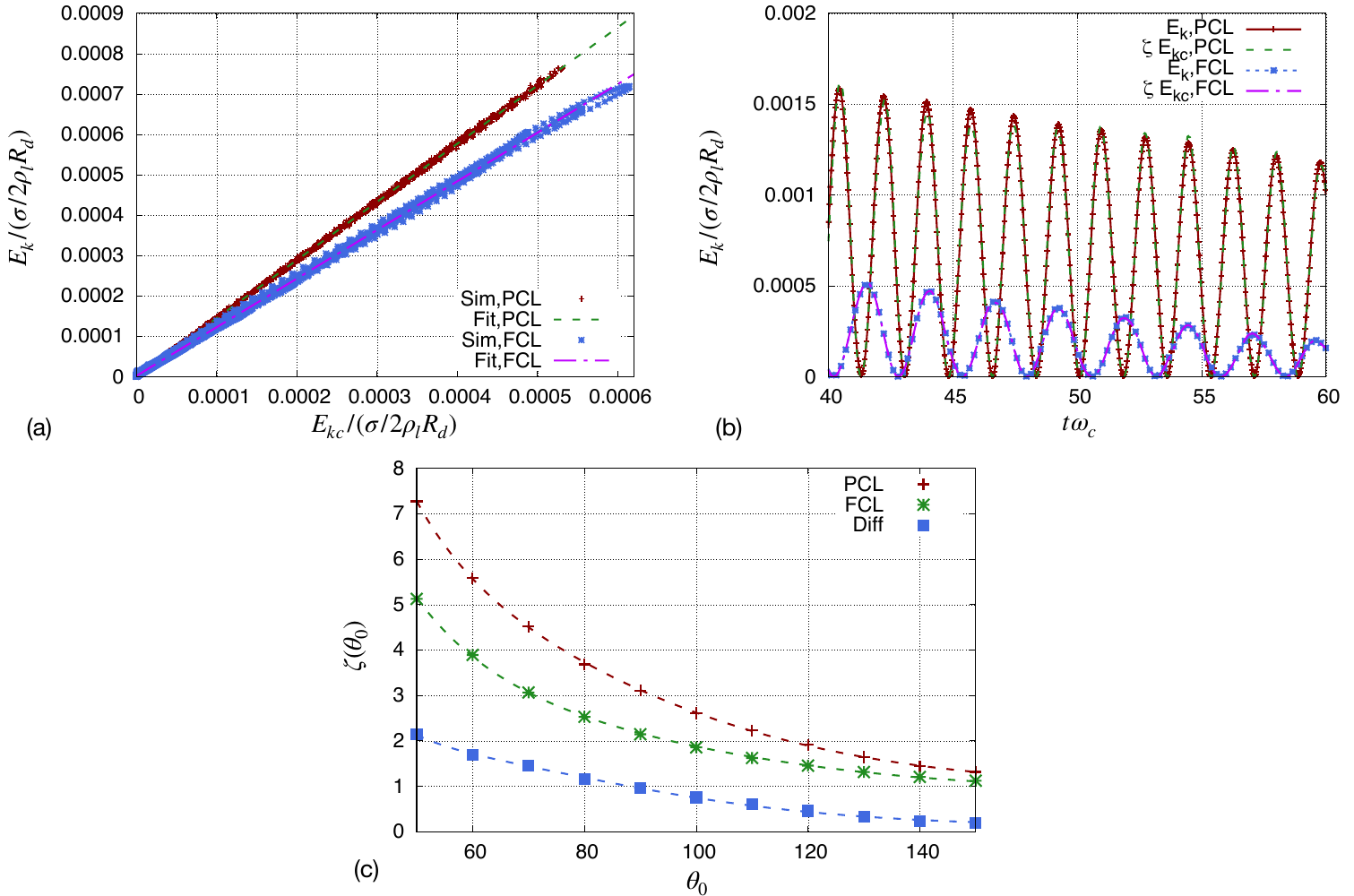}
 \caption{(a) Variation of $E_k$ as a function of $E_{kc}$ and (b) the temporal evolutions for $E_k$ and $\zeta E_{kc}$ for $\theta_0 = \ang{140}$, $\text{Bo}=0$, and both FCL and PCL conditions. (c) Variations of $\zeta$ for PCL and FCL conditions and the difference between the two as a function of contact angle $\theta_0$ for $\text{Bo}=0$. The dashed lines in (c) are fitting correlations. }
 \label{fig:ca140Bo0EkVSEkc_Both}
\end{figure}

When the drop centroid passes the equilibrium positions (ii) and (iv), the centroid velocity reaches local maximum, and in the mean time the internal flow is also intense. When the drop centroid reaches the local minimum (i) and maximum (iii), $u_c$ becomes zero and the flow around the drop is minimal. This indicates that the two contributions to $E_k$, \ie, the one from the centroid motion ($E_{kc}$) and the one due to the shape-oscillation-induced internal flow ($E_k-E_{kc}$) are in phase. Correspondingly, the temporal evolutions of $E_k$ and $E_{kc}$ are also in phase. 
This conclusion is further confirmed in Fig.\ \ref{fig:ca140Bo0EkVSEkc_Both}(a)), where $E_k$ is plotted as a function of $E_{kc}$  for $\theta_0=\ang{140}$ and both FCL and PCL conditions. It is observed that, for both cases, $E_k$ varies approximately linearly with $E_{kc}$. This interesting feature allows us to make an approximation of  $E_k$ as 
\begin{equation}
	E_k \approx \zeta E_{k,c}\, ,
 \label{eq:EK2}
\end{equation}
where $\zeta$ is the kinetic energy correction factor, which is time independent. The value of $\zeta$ can be obtained by fitting the simulation results of $E_k$ vs $E_{kc}$ (see Fig. \ref{fig:ca140Bo0EkVSEkc_Both}(a)). It can been seen that the PCL case exhibits a steeper slope than the FCL case, thus the $\zeta$ value is greater. The temporal evolutions of $E_k$ and the approximation $\zeta E_{kc}$ are plotted in Fig. \ref{fig:ca140Bo0EkVSEkc_Both}(b), which affirms that $\zeta E_{kc}$ agrees well with $E_k$ for all time for both FCL and PCL condtions. Here we only show the results for $\theta_0=\ang{140}$ as an example. The approximation Eq.\ \eqr{EK2} is valid for all $\theta_0$. {The kinetic energy correction factor and the approximation Eq.~\eqr{EK2} are} useful to develop theoretical model to predict the oscillation frequency as shown in previous study \cite{Sakakeeny_2020a}. 

The variation of $\zeta$ as a function of $\theta_0$ for $\text{Bo}=0$ is shown in Fig.\ \ref{fig:ca140Bo0EkVSEkc_Both}(c). Similar to $\omega / \omega_c$, $\zeta$ also decrease with $\theta_0$. For all $\theta_0$, $\zeta$ for PCL is larger that the FCL counterpart. This is again due to the stronger constraint from the surface for PCL. The difference between the values of $\zeta$ for PCL and FCL conditions, \ie, $\zeta_P-\zeta_F$, also decreases with $\theta_0$. As $\theta_0$ approaches $\ang{180}$, both $\zeta_P$ and $\zeta_F$ approaches one since there is neither shape deformation for the first mode nor the additional kinetic energy contribution from the oscillation-induced flow. As a result, $\zeta_P-\zeta_F$ will reach zero. 

Similar to the oscillation frequency,  correlations are also made for $\zeta$ for the PCL and FCL conditions as a function of $\theta_0$ in a similar form: 
\begin{align}
	\log(\zeta(\theta_0)) = e_0+e_1 (1+\cos\theta_0) +\left[\exp\left(\frac{(1+\cos\theta_0) ^{e_2} }{e_3} \right)-1\right]\,.
	\label{eq:zeta_fit}
\end{align}
The fitted constants are $[e_0, e_1, e_2, e_3]=[0,\, 0.753,\, 5.87,\, 54.6]$ for FCL and  $[0.14,\,1.00,\, 9.54,\, 599]$ for PCL. The fitting correlations are plotted in Fig.~\ref{fig:ca140Bo0EkVSEkc_Both}(c) are found to well represent the simulation results.

\subsection{Viscous damping of oscillation}
Due to the viscous effect, fluid motion induced by shape oscillation will dissipate the energy provided by the initial excitation. As a result, the oscillation amplitude will decay over time. In the linear regime, the oscillation amplitude $A$ follows the exponential function in time, 
\begin{equation}
	A(t) = A_0 e^{-\beta t} ,
 \label{eq:Fit_Peaks}
\end{equation}
where $A_0$ is the initial amplitude.
For a free drop, the damping rate normalized by the viscous frequency, \ie, $\beta/\omega_v$  is a function of the mode number, as indicated in Eq.\ \eqr{Beta_Lamb}. For a supported drop, the damping rate will also be influenced by the contact angle and the contact line mobility. The decay of the oscillation amplitude for the present problem mainly reflects the damping rate of the dominant $n=1$ mode. 

\begin{figure}
 \centering
 \includegraphics[width=\linewidth]{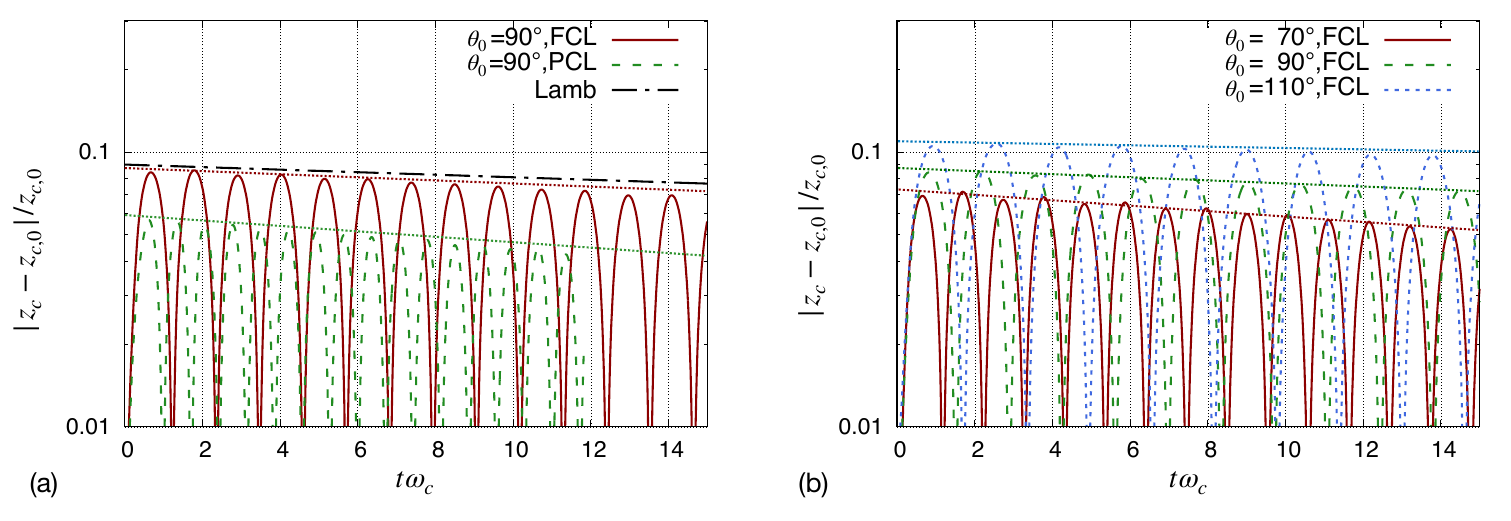}
 \caption{Decay of oscillation amplitude for (a) different contact line mobility and (b) different contact angles for $\text{Bo}=0$. The Lamb damping rate (Eq.\ \eqr{Beta_Lamb}) for the $n=2$ mode is shown for comparison.}
 \label{fig:Damping}
\end{figure}

The temporal evolution of $|z_c-z_{c,0}|/z_c$, for $\theta_0=\ang{90}$, is plotted in Fig.\ \ref{fig:Damping}(a) for both FCL and PCL. As discussed above, the $n=1$ mode of the supported drop with FCL and $\theta_0=\ang{90}$ is similar to the $n=2$ mode of the free drop with twice the size. Therefore, the damping rate $\beta$ is expected to be the same as the $\beta_\text{Lamb}$ for $n=2$, as given in Eq.\ \eqr{Beta_Lamb}. In Fig.\ {\ref{fig:Damping}(a)}, it is can be observed that the oscillating amplitude decay for FCL and $\theta_0=\ang{90}$ agrees very well with the Lamb's prediction. When the contact line is pinned, the damping rate increases slightly. {
The oscillation amplitude damping is due to viscous dissipation of kinetic energy. For FCL, since the slip boundary condition is invoked on the bottom surface, dissipation is only caused by viscous fluid motion inside the drop. For PCL, additional dissipation is induced by the no-slip boundary condition on the surface and the pinned contact line. Therefore, the oscillation damping rate is higher for the PCL than the FCL cases. }

The viscous damping of oscillation amplitude for different $\theta_0$ are shown in Fig.\ \ref{fig:Damping}(b). The damping rate generally decreases with increasing $\theta_0$. As $\theta_0$ increases, the contact area decreases and the constraint to the drop shape deformation is reduced. In the limit of $\theta_0\to \ang{180}$, the drop will not deform due to the $n=1$ mode. As a result, there will be no viscous dissipation  due to the shape oscillation and $\beta \to 0$. 

\subsection{Effect of gravitational Bond number}
\label{sec:finiteBo}
The results discussed so far are only for $\text{Bo}=0$, which represents the oscillation dynamics of supported drops in {a} zero-gravity environment. The results can serve as approximations for tiny drops with very small Bo. However, for the droplet considered ($R_d=5$ mm), when gravity is present, the equilibrium shape of the drop will significantly deviate from the spherical cap. In the present study, we allow $g$ and $\text{Bo}$ to vary from negative to positive values.  For $\text{Bo}>0$, the drop will be flattened, while for $\text{Bo}<0$, the drop will be elongated. The addition of the hydrostatic pressure will also change the pressure balance at the drop surface. The radius of curvature of the drop {at equilibrium state} will not be constant as for Bo=0. The gravity effect is shown to influence both the oscillation frequency and the kinetic energy of the supported drop, though its effect on the viscous damping rate seems to be very minor. 

\begin{figure}[tbp]
 \centering
 \includegraphics[width=\textwidth]{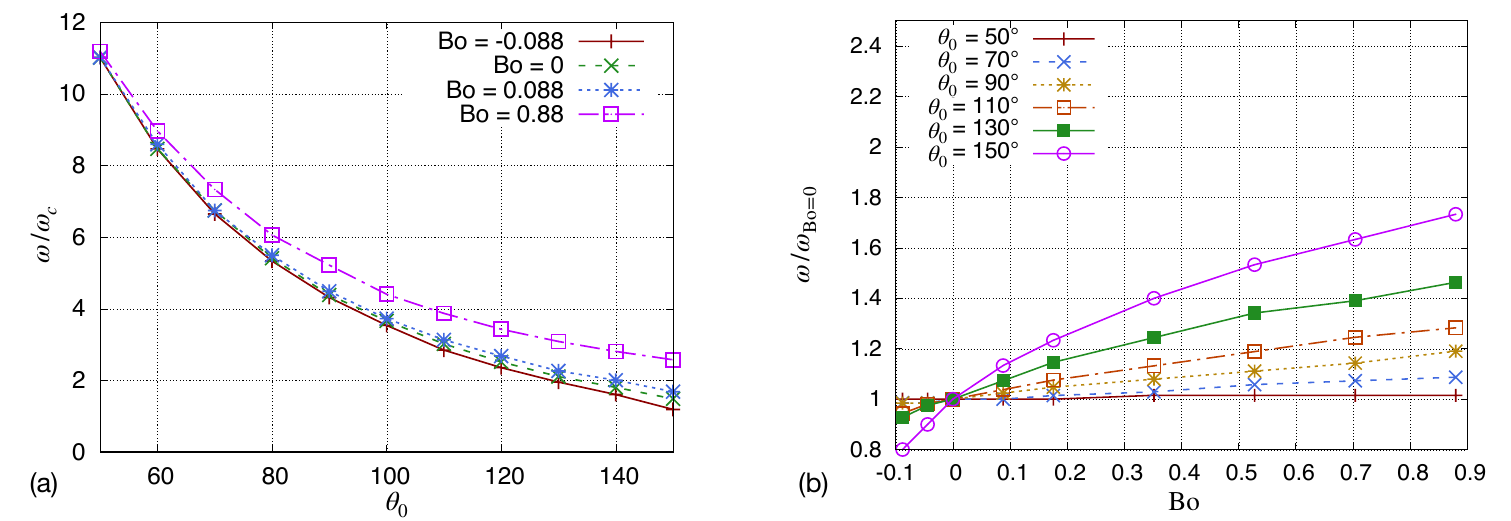}
 \caption{Variation of the supported drop first-mode oscillation frequency as functions of (a) $\theta_0$ and (b) $\text{Bo}$ for the PCL condition.}
 \label{fig:PinnedWWc_combined}
\end{figure}

The simulation results of the first-mode frequencies for supported drops for PCL and different $\theta_0$ and Bo are shown in Fig.\ \ref{fig:PinnedWWc_combined}. It is observed from Fig.\ \ref{fig:PinnedWWc_combined}(a) that for all Bo, the decreasing trends of $\omega / \omega_c$ over $\theta_0$ are similar. Furthermore, the oscillation frequencies increases monotonically with $\text{Bo}$ from negative to positive values for all $\theta_0$. In other words, the oscillation frequency increases with gravity for sessile drops, while for pendant drops, the frequency will decrease due to the gravity effect. The trend of variation of the oscillation frequency is consistent with former observations for both pendant and sessile drops \cite{Basaran_1994a, Sakakeeny_2020a}. Comparing the cases with zero and full gravity, \ie, $\text{Bo}=0$ and 0.88, it is seen that the rise of frequency due to the gravity effect increases with $\theta_0$. 

To better illustrate the change in frequency due to the gravity effect, the oscillation frequency is normalized by that for $\text{Bo}=0$ and is plotted as a function of Bo in Fig.\ \ref{fig:PinnedWWc_combined}(b). Again, the monotonic increase of $\omega / \omega_c$ over Bo from -0.088 to 0.88 can be clearly seen. The rate of increase, indicated by the slopes of the curves, generally decreases with Bo. 

It is further observed that, the rate of change of $\omega / \omega_{Bo=0}$ over $\text{Bo}$ is more significant for large $\theta_0$. For $\theta_0=\ang{150}$, $\omega/\omega_{Bo}$ increases about 72\% when $\text{Bo}$ increases from 0 to 0.088. In contrast, with the same increase of $\text{Bo}$, $\omega/\omega_{Bo=0}$ for $\theta_0 = \ang{50}$ only increases less than 2\%. It is also worth noting that for $\theta_0=\ang{150}$, the frequency decreases quite rapidly with Bo when $\text{Bo}<0$. The results indicate that the oscillation frequency for drops supported by hydrophobic/super-hydrophobic surfaces can be very sensitive to the change of $\text{Bo}$. 

{The variation of the oscillation frequency with $\theta_0$ and Bo can be explained by the inviscid theoretical model developed in our previous study \cite{Sakakeeny_2020a}, in which the first-mode oscillation can be modeled as a mass-spring harmonic oscillator, 
\begin{equation}
	 k (z_c - z_{c,1}) + m \frac{d^2  (z_c - z_{c,1})}{dt^2} = 0,
 \label{eq:HarmonicOscillator}
\end{equation}
where $k$ and $m$ are the effective spring constant and drop mass, while $z_{c,1}$ represents the equilibrium centroid location for finite Bo. For $\text{Bo=0}$, $z_{c,1}=z_{c,0}$. Here, the restoring force, $k (z_c - z_{c,1})$, is mainly due to surface tension. It was shown that $k\sim \eta$, where $\eta$ is a parameter that characterizes the increase of the drop surface area as the centroid deviates from the equilibrium position, namely $(S-S_1)/S_0=\eta((z_c - z_{c,1})/R_0)^2$, where $S_0$ and $S_1$ are the equilibrium drop surface area for zero and finite Bo. As a result, the oscillation frequency 
\begin{equation}
	\omega^2=k/m\sim \eta\, .
\end{equation}
The $S$-$z_c$ curve and $\eta$ for a given $\theta_0$ and Bo can be estimated by the equilibrium drop theory, see Ref.\ \cite{Sakakeeny_2020a} for details. }

{The equilibrium drop theory indicates that $\eta$ monotonically increases with Bo for all $\theta_0$. In other words, when Bo increases, the drop equilibrium shape deviates from the spherical cap, and the increase of surface area $(S-S_1)$ for a given centroid deviation $(z_c-z_{c,1})$ becomes higher. As a consequence, the restoring force increases, and thus too the oscillation frequency increase. 
Similarly, the rate of increase of $\eta$ with Bo also increases with $\theta_0$. For supported drops with large $\theta_0$,  the surface area increase is more ``responsive" to the centroid deviation and the change of Bo, due to the smaller contact area and constraint from the surface. Therefore, the difference between the surface area increments for the same $(z_c-z_{c,1})$ for zero and finite Bo,  namely $(S-S_1)-(S-S_1)_{\text{Bo=1}}$, is higher for larger $\theta_0$. Correspondingly, the increases in both the restoring force and the frequency are also magnified as $\theta_0$ increases. 
}

\begin{figure}[tbp]
    \centering
    \includegraphics[width=\textwidth]{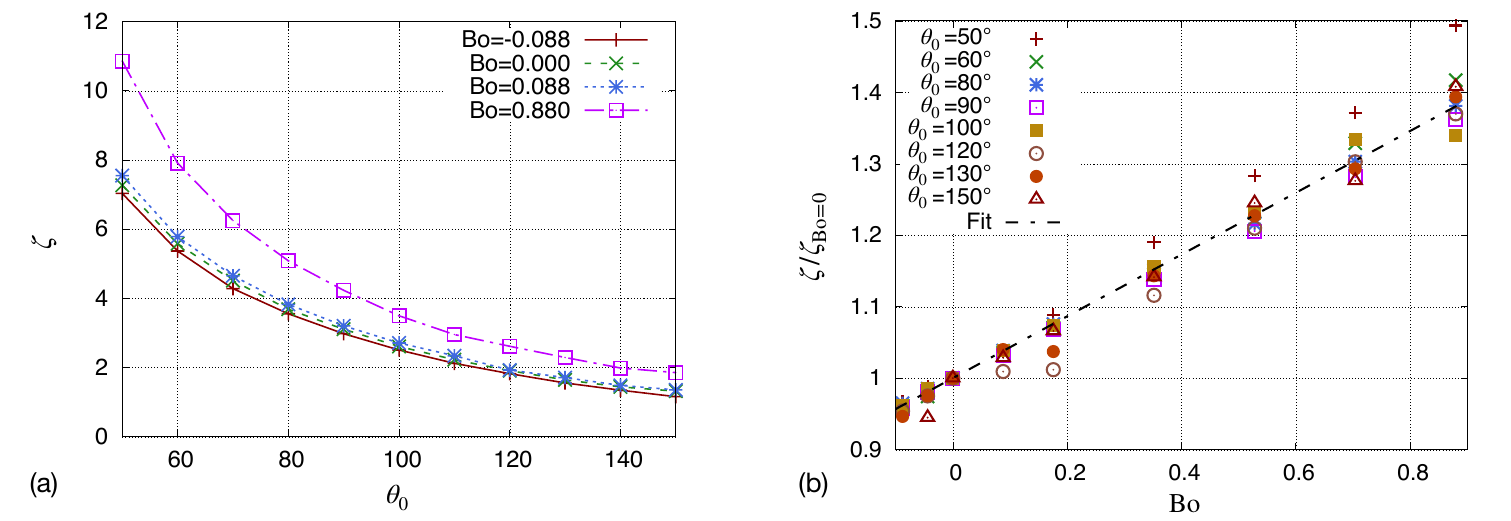}
    \caption{Variation of the kinetic energy correction factor $\zeta$ as functions of (a) $\theta_0$ and (b) $\text{Bo}$ for the PCL condition.}
    \label{fig:Zeta_Vals_Pinned}%
\end{figure}

The gravity effect also modifies the fluid flow induced by shape oscillation and the drop kinetic energy correction factor $\zeta$. The simulation results of $\zeta$ for different $\theta_0$ and $\text{Bo}$ and for the PCL condition are shown in Fig.\ \ref{fig:Zeta_Vals_Pinned}. Similar to the oscillation frequency, $\zeta$ decreases with $\theta_0$ for all $\text{Bo}$ and increases with $\text{Bo}$ for all $\theta_0$. Yet {unlike} $\omega$, the increase of $\zeta$ due to the rise of Bo is more profound for smaller $\theta_0$. Since the value of $\zeta$ also decreases {as} $\theta_0$ decreases, it ends up that the normalized results, \ie, $\zeta/\zeta_{Bo=0}$, for different $\theta_0$ collapse approximately, see Fig.\ \ref{fig:Zeta_Vals_Pinned}(b), and can be fit via a linear function of $\text{Bo}$ as
\begin{equation}
	\zeta(\theta_0,\text{Bo}) = \zeta_{Bo=0}(\theta_0)(1+\alpha \text{Bo})\, .
	\label{eq:zeta_general_fcn}
\end{equation}
For the PCL results for all $\theta_0$ and Bo, the fitting yields $\alpha= 0.432$. Similar scaling behavior has been observed for the FCL condition \cite{Sakakeeny_2020a}, where $\alpha=0.358$. Therefore, the increase of the kinetic energy correction factor over Bo is faster when the contact line changes from the FCL to the PCL condition.

\begin{figure}[tbp]
    \centering
    \includegraphics[width=\textwidth]{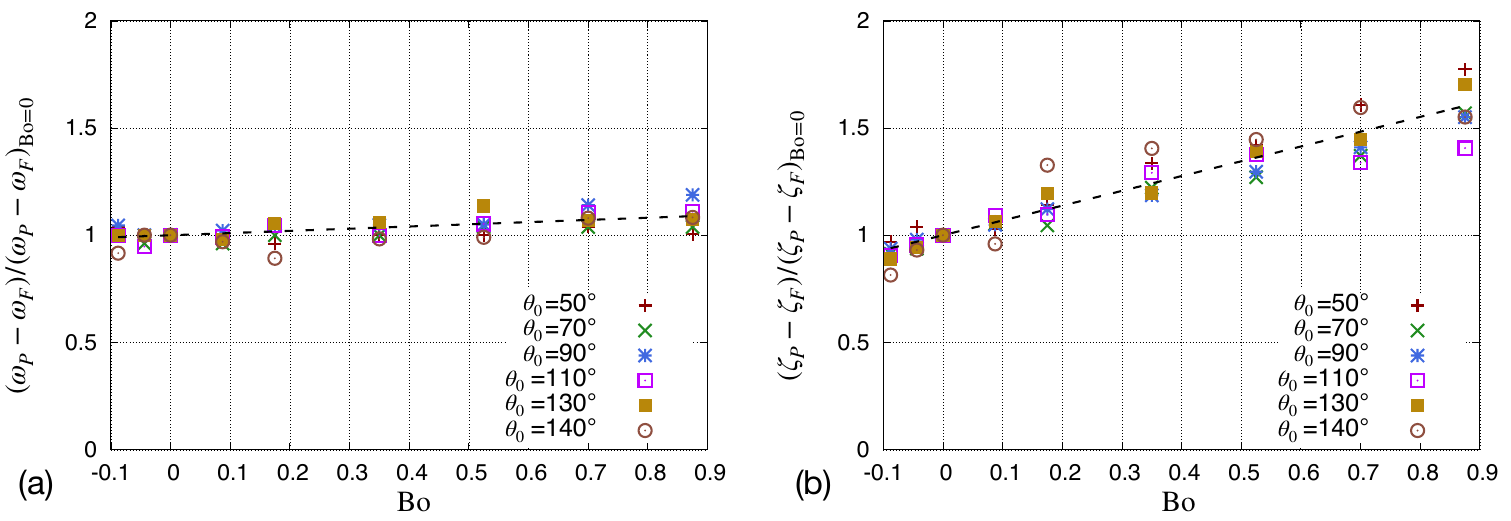}
    \caption{The differences between the values corresponding to the PCL and FCL conditions for (a) $\omega$ and (b) $\zeta$ as functions of $\theta_0$ and Bo.}
    \label{fig:Diff_Bo}%
\end{figure}

For all $\theta_0$ and $\text{Bo}$, the oscillation frequency $\omega/\omega_c$ and the kinetic energy correction factor $\zeta$ for PCL are always larger than their FCL counterparts. The difference between the PCL and FCL values of $\omega/\omega_c$ and $\zeta$, normalized by the difference at Bo=0, \ie, $(\omega_P-\omega_F)/(\omega_P-\omega_F)_{\text{Bo}=0}$ and  $(\zeta_P-\zeta_F)/(\zeta_P-\zeta_F)_{\text{Bo}=0}$, are plotted in Fig.~\ref{fig:Diff_Bo}. For both variables, the results for different $\theta_0$ collapse approximately. It is further observed that $(\omega_P-\omega_F)/(\omega_P-\omega_F)_{\text{Bo}=0}$ varies little with Bo. The collapsed results approximately follow a linear function passing through the point (0,1) with a very small slope (about 0.10). This implies that, though the frequencies $\omega_P$ and $\omega_F$ increase over $\text{Bo}$, the difference between the two actually changes little. As a result, the difference between oscillation frequencies for the PCL and FCL at Bo=0, $(\omega_P-\omega_F)_\text{Bo}=0$, see  Fig.~\ref{fig:xc_IC1_Bo0_theta_combined}(c), is a good approximation for non-zero Bo cases. On the other hand, the results for $(\zeta_P-\zeta_F)/(\zeta_P-\zeta_F)_{\text{Bo}=0}$ for different $\theta_0$ also approximately collapse and agree with a linear function, but the slope is bigger than that for the frequency, \ie, about 0.690. As a result, the difference between the PCL and FCL values of $\zeta$, \ie, $(\zeta_P-\zeta_F)_{\text{Bo}=0}$ as shown in Fig.~\ref{fig:ca140Bo0EkVSEkc_Both}(c), needs to be corrected using the results shown in Fig.~\ref{fig:Diff_Bo}(b) to represent the non-zero Bo cases. 

\section{Conclusions}
\label{sec:conclusions}
The axisymmetric natural oscillations of a liquid drop supported by a flat surface have been studied by direct numerical simulation. The parameters, including the equilibrium contact angle ($\theta_0$) and the gravitational Bond number (Bo), are varied to systematically investigate their effects on the oscillation frequency and the induced flow around the drop. The two asymptotic limits of contact line {hysteresis} and mobility, \ie, the pinned and free contact line conditions, are considered to investigate the effect of contact line mobility on oscillation. For the pinned contact line (PCL) condition, the drop contact angle can vary freely, but the contact line cannot move. For the free contact line (FCL) condition, the drop contact angle is fixed, while the contact line is allowed to move freely. The results of oscillation frequencies for these two limiting cases also serve as the upper and lower bounds for general contact line conditions. In total, over 198 simulation cases were performed to study a wide range of {equilibrium} contact angles ($\ang{50} \leq \theta_0 \leq \ang{150}$) and Bond numbers ($-0.088 \leq \text{Bo} \leq 0.88$) for both the FCL and PCL conditions. The negative and positive Bo represent the pendant and sessile drops, respectively. 

The drop oscillation is initiated by changing the gravity for a short period of time. The first oscillation mode due to the drop centroid translation is observed to dominate the excited oscillations. The oscillation frequency $\omega$ scales with the capillary frequency $\omega_c$, and the normalized frequency $\omega/\omega_c$ decreases with $\theta_0$.  Remarkable agreement between the simulation results with the inviscid theory of Bostwick and Steen \cite{Bostwick_2014a} is achieved, which validates the present simulations. The shape oscillations induce flows within the drop that contributes to the kinetic energy of the drop. The kinetic energy correction factor $\zeta$ is defined as the ratio between the total kinetic energy of the drop and that for the bulk motion. Similar to $\omega$, $\zeta$ also decreases with $\theta_0$. The viscous damping rate $\beta$ of the oscillation amplitude is also observed to decrease with $\theta_0$. 

When $\text{Bo}$ increases from -0.088 to 0.88, both $\omega/\omega_c$ and $\zeta$ increase. The increase in $\omega/\omega_c$ due to the rise of gravity becomes more profound for larger $\theta_0$, indicating that the drop oscillation frequency for hydrophobic/superhydrophobic surfaces can be quite sensitive to the gravity effect. In contrast, the increase of $\zeta$ due to gravity is more significant for small $\theta_0$. Furthermore, the results of $\zeta$ for different $\theta_0$ collapse to a linear function if they are normalized by the values at zero Bo. For all $\theta_0$ and $\text{Bo}$, the values of $\omega/\omega_c$ and $\zeta$ for PCL  are always greater than their respective FCL values. The difference between the frequencies for FCL and PCL, $\omega_P-\omega_F$ for different $\theta_0$ scales with the counterpart for $\text{Bo}=0$, and the normalized difference, $(\omega_P-\omega_F)/(\omega_P-\omega_F)_{\text{Bo}=0}$, varies little with $\text{Bo}$.

\section*{Acknowledgement}
This work was supported by the startup fund at Baylor University and the National Science Foundation (1853193, 1942324). The Baylor High Performance and Research Computing Services (HPRCS) have provided the computational resources that have contributed to the research results reported in this paper. The authors also acknowledge Dr.~St\'ephane Popinet for the contribution to the development of the Basilisk code. 

\section*{Data Availability}
The data that support the findings of this study are available from the corresponding author upon reasonable request.

%

\end{document}